# Impact Factor: outdated artefact or stepping-stone to journal certification?

Jerome K Vanclay

Southern Cross University

PO Box 157, Lismore NSW 2480, Australia

JVanclay@scu.edu.au

**Abstract**

A review of Garfield's journal impact factor and its specific implementation as the Thomson Reuters Impact Factor reveals several weaknesses in this commonly-used indicator of journal standing. Key limitations include the mismatch between citing and cited documents, the deceptive display of three decimals that belies the real precision, and the absence of confidence intervals. These are minor issues that are easily amended and should be corrected, but more substantive improvements are needed. There are indications that the scientific community seeks and needs better certification of journal procedures to improve the quality of published science. Comprehensive certification of editorial and review procedures could help ensure adequate procedures to detect duplicate and fraudulent submissions.

## Introduction

Increasing scrutiny of the Thomson Reuters Journal Impact Factor casts some doubt on its claim of a "*systematic, objective means to critically evaluate the world's leading journals, with quantifiable, statistical information based on citation data*" (Thomson Reuters 2011). The continued inappropriate use of this indicator, despite serious flaws, invites comparison with phrenology, the out-dated pseudo-science that attempted to infer human behaviour from measurements of skull morphology (Davies 1955). Although popular in the early 19th century, most scientists now recognise that such measurements offered an inaccurate record of morphology and an unreliable indicator of human behaviour. Unlike phrenology, the impact factor (Garfield 1972) has demonstrated utility in informing citation patterns and guiding library purchasing decisions (Althouse et al 2008, Cameron 2005). However, there are increasing concerns that the impact factor is being used inappropriately and in ways not originally envisaged (Garfield 1996, Adler et al 2008). These concerns are becoming a crescendo, as the number of papers has increased exponentially (Figure 1), reflecting the contradiction that editors celebrate any increase in their index, whilst more thoughtful analyses lament the inadequacies of the impact factor and its failure to fully utilize the potential of modern computing and bibliometric sciences. Although fit-for-purpose in the mid 20th century, the impact factor has outlived its usefulness. Has it become, like phrenology, a pseudo-science from a former time? This paper attempts to draw together key conclusions from many extant papers, to illustrate limitations with explicit examples, and to offer recommendations to improve the indicator.

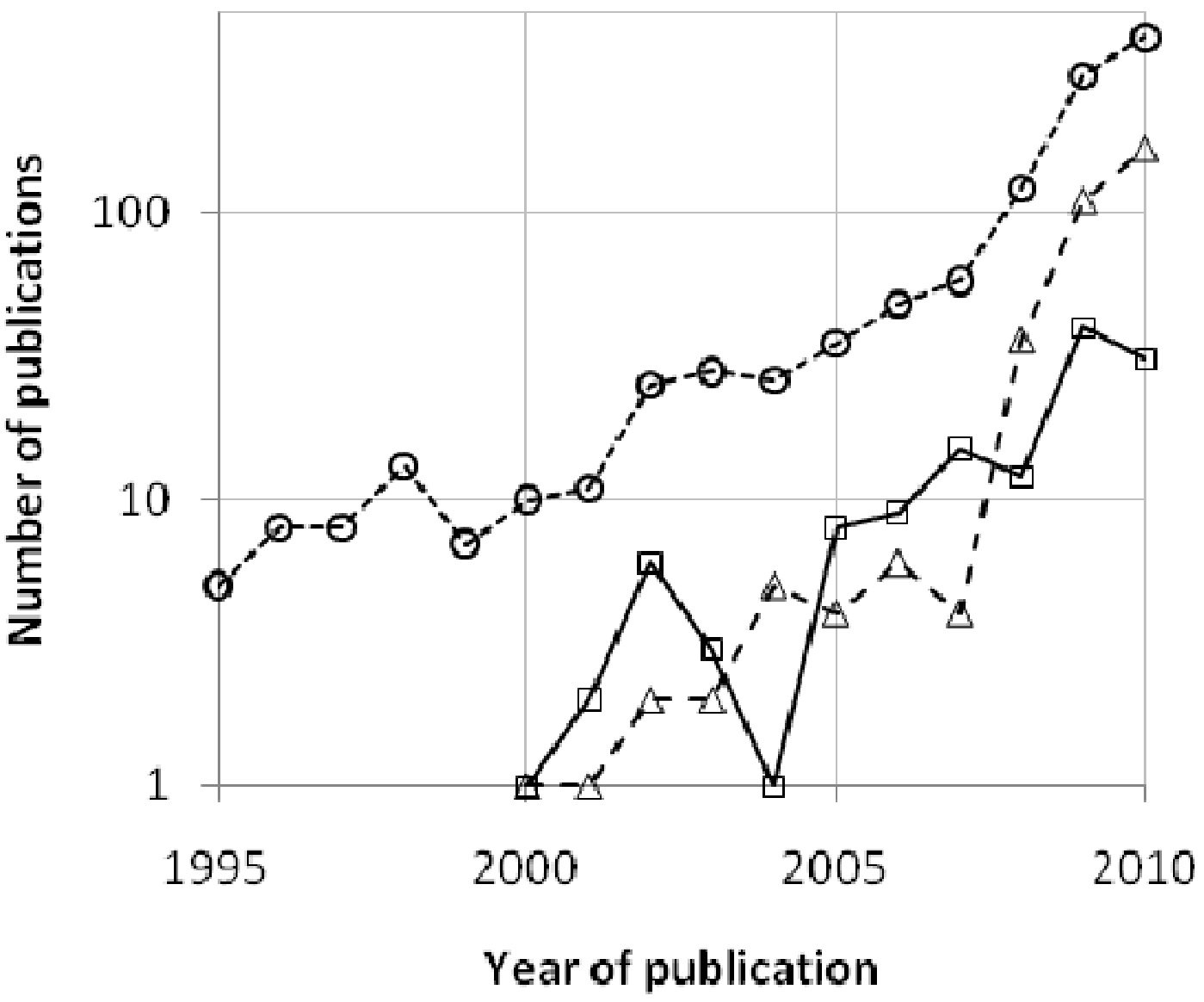

**Figure 1**. Exponential increase in documents found with a Scopus search for 'journal impact factor', showing all documents (dotted line, circles), editorial comment (solid line, squares) and critical documents (dashed line, triangles; with the words 'bias', 'limitation', 'problem', 'manipulate', 'misuse' or 'flaw' in the abstract).

It is appropriate to begin by examining exactly how the impact factor is computed, and how there may be several variants depending on how this computation is made. Thus in this paper, I adopt the term Garfield index to denote the generic concept underpinning the impact factor, and the abbreviation TRIF to denote the specific Journal Impact Factor published by Thomson Reuters in their Journal Citation Reports (JCR).

## Garfield's impact factor

In 1972, Eugene Garfield (1972, 1995, 2006) proposed an impact factor, based on the mean number citations to articles published in the two preceding years. To avoid confusion with particular commercial implementations, let's call this general concept the Garfield index, and define it as the ratio of citations received divided by the potential number of cited documents. Before exploring the variants of this index, it is appropriate to recognise that technological challenges shaped some compromises made in the 1970s (Garfield 1972), and to note that nothing in the TRIF challenges the more advanced computational capability available today. There may be several variants of Garfield's index, reflecting how both the numerator and the denominator of this ratio are derived. In principle, the numerator of this ratio is simply the number of citations in year *y* to articles published in years *y-1* and *y-2*, but there are a number of ways to deal with ambiguous citations (Figure 2).

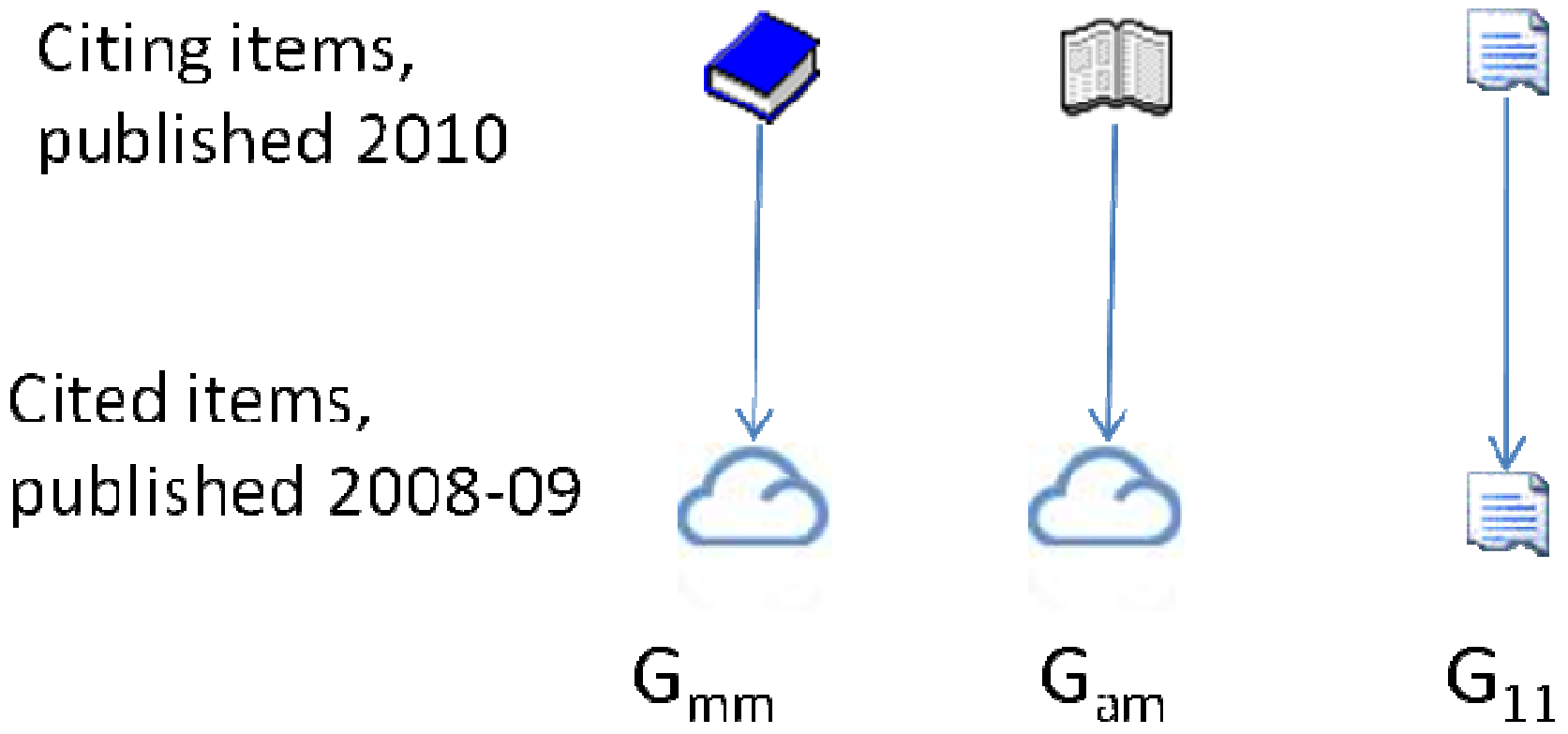

**Figure 2**. Variants of the numerator forming the Garfield index.

The simplest form of the Garfield index relies on the many-to-many numerator *Gmm* (Figure 2). Its simplicity arises from the ease of extracting from a database a count of all references (in year *y*) to journal *J* in year *y-1* or *y-2*. It is fault-tolerant, because it includes all citations matching journal title and year, even if there are other errors in the citation. Figure 2 deliberately shows a cloud to signify the uncertainty associated with the cited article: there may or may not be an article corresponding to the citation as no cited-side checking is done to ensure the integrity of the citation. The Thomson Reuters Impact Factor (TRIF, Garfield 1994) is a particular implementation of the *Gmm* index drawn from the Web of Science (WoS) database and published in the Journal Citation Reports (JCR).

A more discerning form of the numerator *Gam* takes into account the nature of the citing document and restricts the count of citations to scientific articles (i.e., excluding editorial comment), thus eliminating some spurious content (e.g., by Google Scholar in indexing 'Indices to Volume 58' by the author Area Index) as well as mischievous editorials promulgating self-citations designed to inflate the index (e.g., Hernan 2009). The preferable form is the more rigorous $G_{11}$, which checks for the existence of the cited document, and ensures the validity of the one-to-one link defined by the citation. Database professionals would use different terminology, but it is useful to retain the $G_{11}$ notation to emphasise that each citing paper must link explicitly with each cited paper in a one-to-one link. To clarify, *Gmm* involves minimal checking of citations, *Gam* involves citing-side checking, and $G_{11}$ involves both citing-side and cited-side checking. This quality control implicit in the $G_{11}$ variant solves many of the problems of the common *Gmm* form of the numerator, but creates new challenges (if an author mis-quotes a reference - even with a minor typographic error - the database cannot create a link: should the incorrect reference remain, pointing nowhere, or should an attempt be made to correct the reference and make a link?). However, it sheds transparency on an insidious problem with the conventional fault-tolerant *Gmm* form which does not compare like with like: a compelling advantage of the $G_{11}$ approach is that it enables an analyst to select the relevant material for both the numerator and denominator.

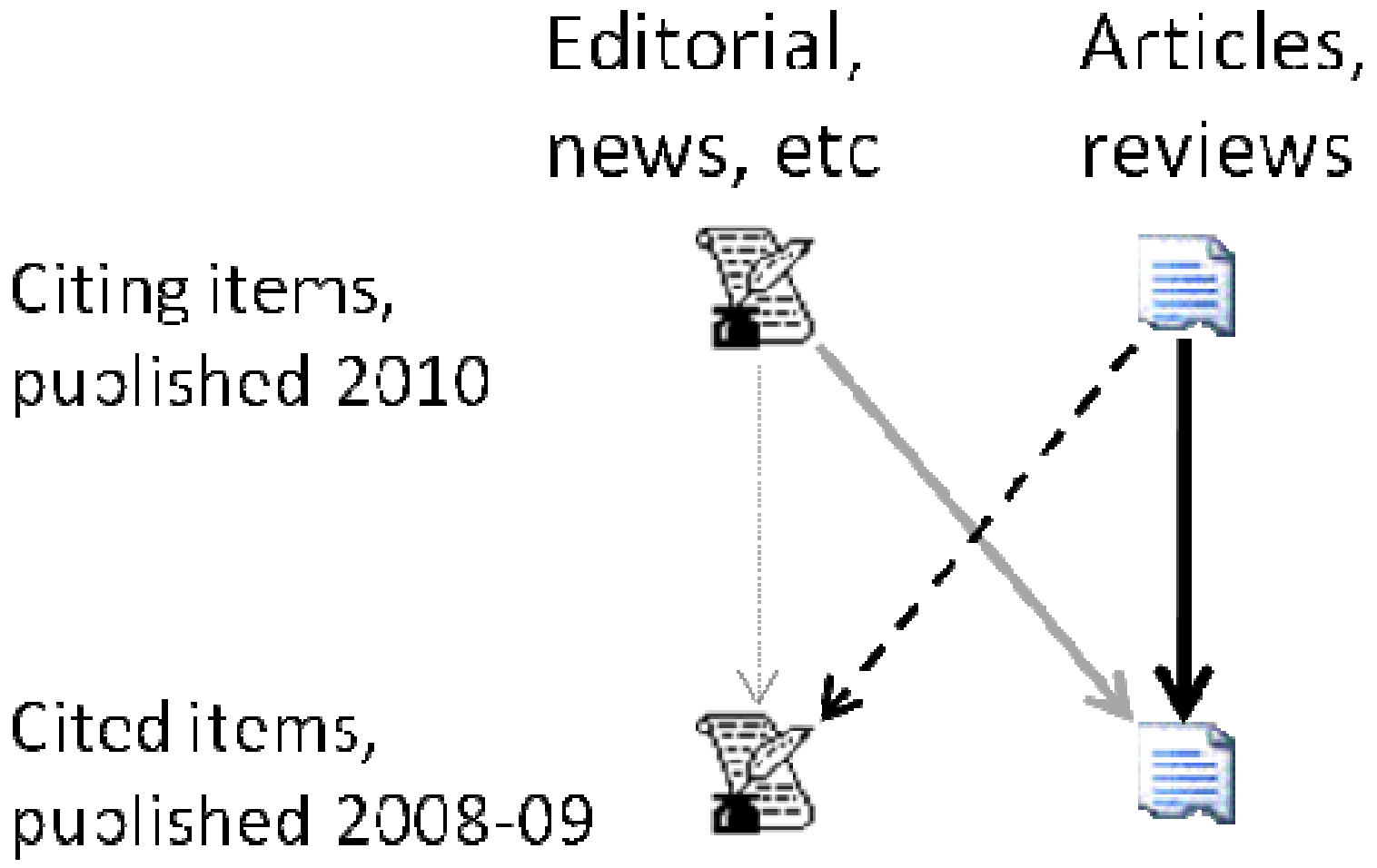

**Figure 3**. Variants of the denominator forming the Garfield index.

Figure 3 illustrates the nature of denominator forming the Garfield index. Most citations are made by articles (including reviews) to earlier articles (solid black line in Fig. 3). Some editorial material may cite articles (items by the editor, and letters to the editor commenting on earlier articles; solid grey line), and this is a way to manipulate the TRIF. There may also be a few references to earlier editorials and letters (dashed black and dotted grey lines), but in many journals, these constitute a small proportion of the total citations received. However, the relative proportions may be surprising: McVeigh and Mann (2009) report several journals in which so-called 'citeable articles' (i.e., articles and reviews) comprise only 20% of the total content. The importance of these distinctions is that while logic might suggest an impact factor should rely on scientific citations (by articles and reviews, black lines), the popular TRIF counts *all* the lines in calculating the numerator (i.e., *Gmm*) and assumed only the solid lines in deriving the denominator (i.e., the porential number of articles and reviews that might receive citations, but excluding editorial material). Thus the TRIF uses a ratio of all citations received (years *n-1* and *n-2*) dived by the number of articles potentially available to be cited (black lines only, year *n*). While this was computationally convenient in the past (Archambault & Lariviere 2009) and is harmless in many cases with few editorial citations, it is potentially problematic when editors choose to manipulate their TRIF with wanton self-citations within their own journal (e.g., Rieseberg & Smith 2008, Rieseberg et al 2011). Conversely, editorial improvements to journal layout can also have unintended consequences, such as when, in 1997, the *Lancet* divided its 'Letters' section into 'Correspondence' and 'Research Letters', the latter being peer-reviewed and hence 'citable' for the denominator, the increase in the denominator led to a fall in IF from about 17 to about 12 (Scully and Lodge 2005). If the TRIF is to be taken seriously, it should be revised so that it deals only with $G_{11}$ citations (i.e., both citing-side and cited-side checking) by articles and reviews, to articles and reviews (i.e., in Figure 3, solid black line only).

A third aspect of citation analysis is self-citation: citations to articles may originate from within a journal, or from other journals. In general, most citations originate from other journals, but the proportion of self-citation varies with discipline and journal. Generally, self-citation rates for most journals remain below 20% (McVeigh 2004), but may be somewhat higher for specialist and national journals (Fassoulaki et al 2000). For instance, in 2010, *Australian Forestry* had a self-citation rate of about 25% (Table 1). *Australian Forestry* was chosen deliberately for this analysis because it is a small journal familiar to the author; both necessary criteria because (with the present versions of WoS and Scopus) the reliable compilation of Table 1 required manual inspection of each citation.

**Table 1**. Origins of citations in 2010 to items published in *Australian Forestry* during 2008-09

| Source | Web of Science | | | Scopus | | |
|---|---|---|---|---|---|---|
| | Article | Letter | Total | Article | Letter | Total |
| *Australian Forestry* | 10 | 1 | 11† | 11 | 1 | 12 |
| *N.Z. J. Forestry Science* | | | | 5 | | 5 |
| *Int. J. Appl. Earth Obs. and Geoinf.* | 4 | | 4 | 4 | | 4 |
| *Forest Ecology and Management* | 3 | | 3 | 3 | | 3 |
| *Australian Journal of Entomology* | 3 | | 3 | 3 | | 3 |
| Other journals | 23 | | 24 | 22 | | 22 |
| Faulty references | 1 | | | | | |
| Total | 44 | 1 | 45† | 48 | 1 | 49 |

† These data from a cited reference search in WoS; JCR reports 12 self-cites amongst a total of 46 citations, but it is not possible to reproduce this from the WoS.

Table 1 also reveals a fourth aspect about the Garfield index: the values obtained for any journal will depend on the database, because each provider utilises a different collection of source material. It is evident in Table 1 that Scopus scans the *New Zealand Journal of Forestry Science*, while the WoS does not. In addition, WoS noticed citations (unseen by Scopus) to *Australian Forestry* in *Methods in Ecology and Evolution*, in *Forest Systems*, and in *Geoderma*. Conversely, Scopus noticed citations unseen by WoS in *Papers and Proceedings of the Royal Society of Tasmania*, and in *New Zealand Journal of Forestry*. A third provider, Google Scholar, found 33 citations including 22 of those represented in Scopus, plus additional citations in *CAB Reviews: Perspectives in Agriculture, Veterinary Science, Nutrition and Natural Resources*, and amongst grey literature not scanned by the commercial providers (conference proceedings, technical reports, etc.).

The TRIF is displayed to three decimals by convention, apparently to create a unique ranking and to minimize the number of tied places (Garfield 2006), but this is a misleading practice. Consider an informative and familiar analogy, the body mass of a group of people. While it is possible to obtain scales that display to the nearest gram (e.g., 65.432 kg), such detail is irrelevant, because the value displayed may depend on the time of day (time since last meal or bathroom visit), the surface supporting the scales (Sample 2002), the mechanism, and the manufacturer. Thus in most cases, people deal with human mass to the nearest kilogram, except in carefully controlled studies (Stein et al 2005). So it is with Garfield's index, which may vary from day to day (depending on the last batch of error corrections), and depend on the database used (within WoS: Science Citation Index Expanded, Social Sciences Citation Index, etc; or alternative databases such as Scopus and Google Scholar) and the methodology (*Gmm, Gam* or $G_{11}$ as numerator; articles or all items as denominator). This issue of variability is particularly problematic for journals publishing few articles in any year. For instance, in 2009, *Mediterranean Politics* received 21 citations, indicating an impact of 0.677 and rank 56. In 2010, 19 citations indicated an impact of 0.559 and rank 81. The convention for the TRIF to display three decimal places is illogical and deceitful. Logic indicates not more than one decimal, with many journals tied in equal places.

This paper illustrates that these limitations with the generic Garfield index apply equally to the TRIF, and argues on the basis of literature and new analyses that journal impact should be assessed in ways other than with the current TRIF.

## Literature

There is a vast and rapidly-increasing literature surrounding the TRIF (Figure 1), much of it critical, but the respect afforded the annual update of the JCR (e.g., press releases: Beal 2011, Reller 2011; and editorials: Ho 2011) reflects continuing standing of the TRIF, particularly amongst publishers. Amongst the academics who contribute articles and reviews, there is much more scepticism (e.g., Simons 2008), including some editors and publishers who have called for reform (e.g., Campbell 2008, Patterson 2009). Many concerns have been dealt with repeatedly in earlier reviews (e.g., Braun and Glanzel 1995, Glanzel and Moed 2002, Bensman 2007, 2008, Braun 2007), so the current analysis seeks to offer a brief overview and synthesis to establish a pathway forward. An overview of some aspects is summarised in Figure 4, which displays author-supplied keywords derived from a Web of Science search for ‘impact factor’ and displayed using CiteSpaceII (Chen 2006), one of several software tools offering text analysis (see e.g., Cobo et al 2011).

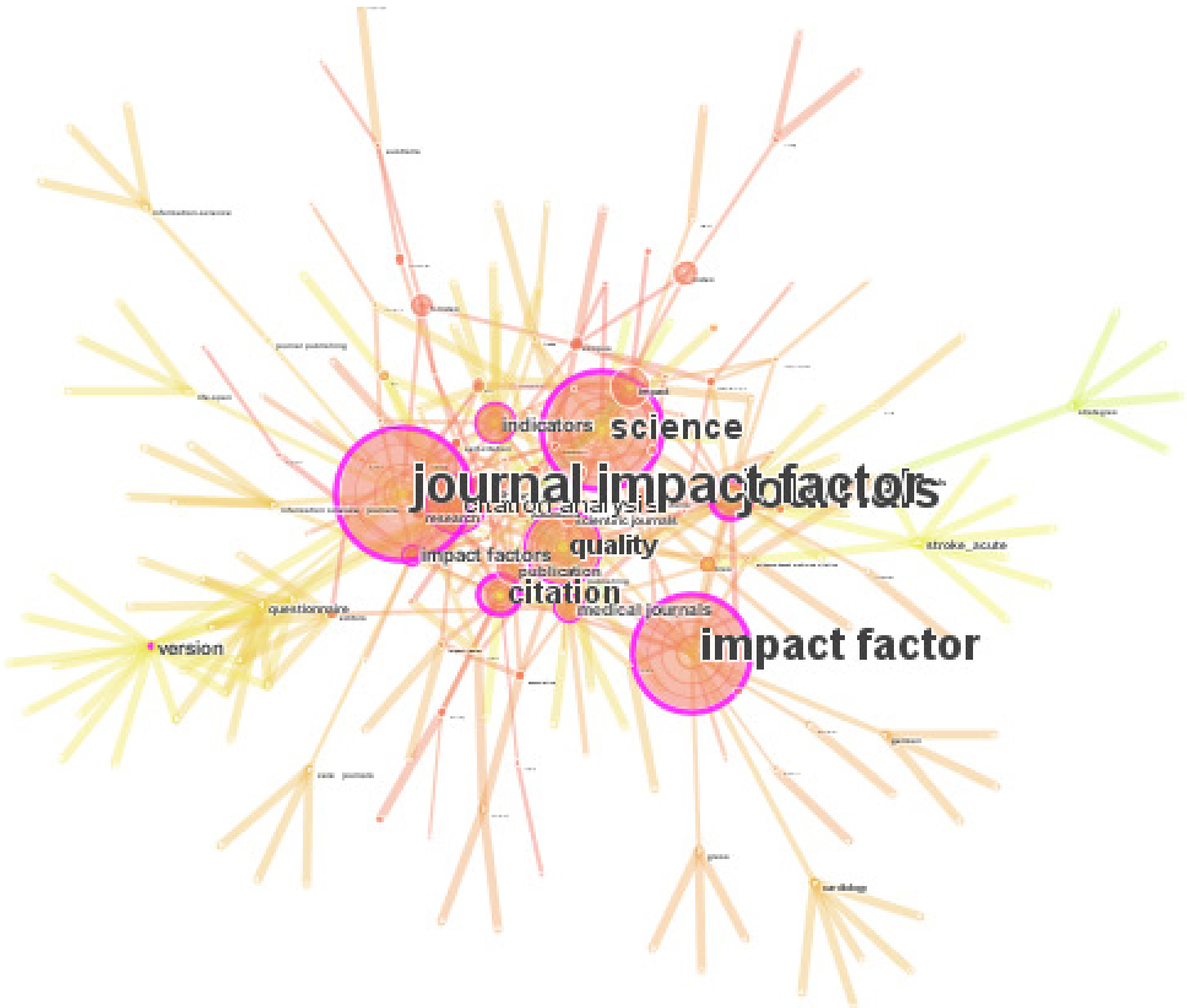

**Figure 4**. CiteSpaceII display of author-supplied keywords amongst 338 articles retrieved with a Web of Science search for ‘impact factor’. ‘Quality’ features prominently in this automated synthesis.

Figure 4 reveals that one of issues often canvassed in conjunction with the journal impact factor is that of ‘quality’. Figure 4 displays a network of co-occurring phrases that were detected automatically, and CiteSpaceII can apply similar techniques to articles, journals and authors to compile article co-citation networks, journal co-citation networks, or author co-citation networks. All of these can be useful to identify important and pivotal material that may not be as conspicuous in a simple list of citations (Jahangiriana et al 2011).

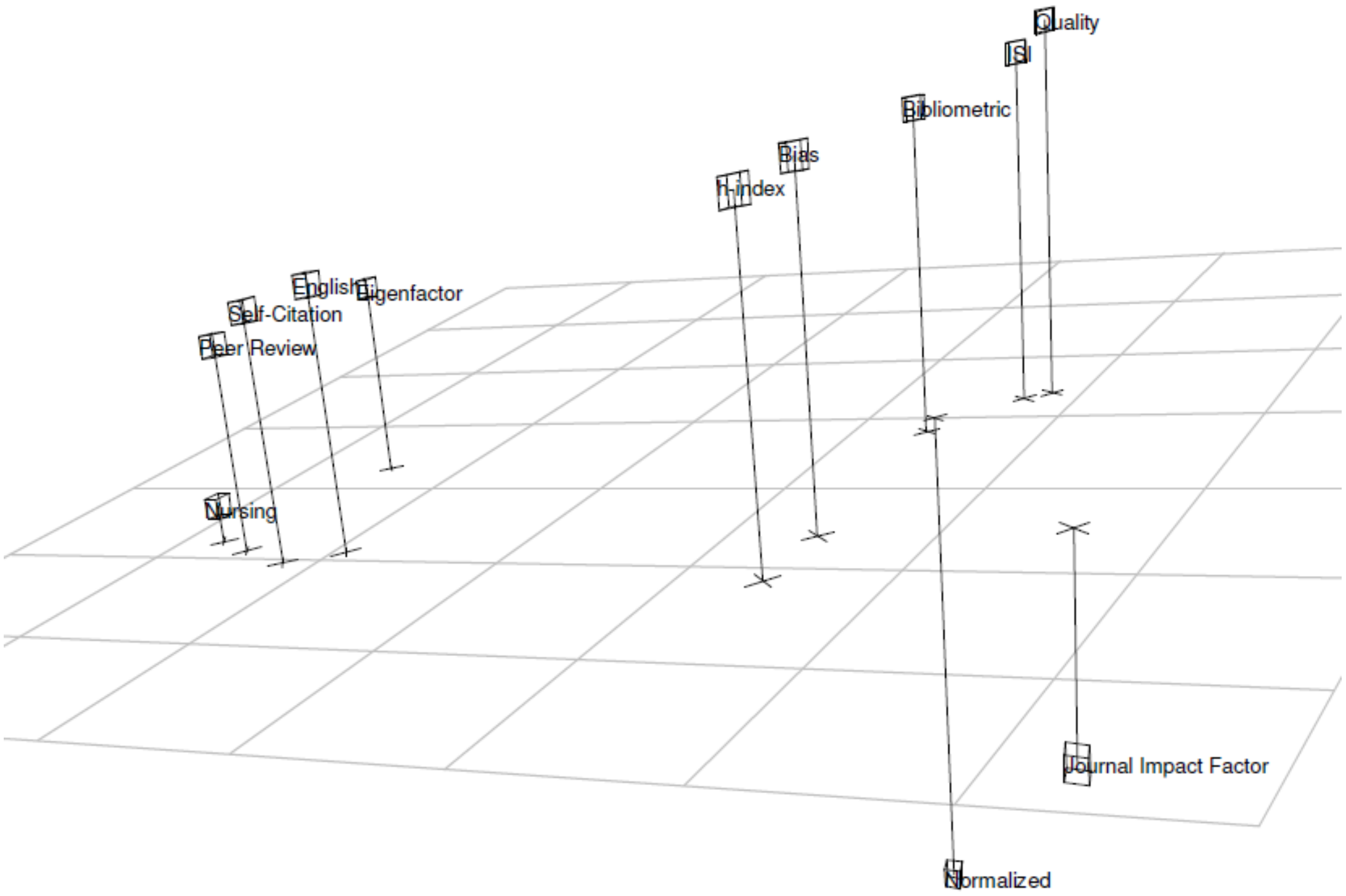

**Figure 5**. Perceptual map compiled with CatPac-Thoughtview from material derived using a Scopus search for 'journal impact factor'.

Other text analysis packages may offer complementary insights: for instance, Figure 5 is a perceptual map compiled from a Scopus search for 'journal impact factor' using CatPac (Woelfel 1993, Jörgensen 2005). Figure 5 makes it easy to note the juxtaposition of 'nursing' and 'peer review', of 'quality' and ISI (predecessor to Thomson Reuters), and of 'normailze' and 'journal impact factor', reflecting some of the issues canvassed in the literature. Catpac (Woelfel 1993, Jörgensen 2005), a text analysis system based on an artificial neural network, and ThoughtView (Woelfel and Woelfel 1997) allow the use more control to 'weed' the image of redundant detail to focus attention on salient features (Figure 5). While such images assist discovery and the formulation of hypotheses, they are ill-suited for hypothesis testing, as the user has ample scope to influence the terms included in or omitted from the display.

These images (Figures 4 and 5) offer an overview of the topic, and illustrate the range of viewpoints expressed and vocabulary used in the literature, and thus help to clarify the trends illustrated in Figure 1. Closer examination reveals that many of the articles found with a simple search merely used (e.g., "The factors related to publication in journals with an impact factor of more than 2 were analysed", Bonillo Perales 2002) or reported the TRIF (e.g., "Our first impact factor", Timuralp 2010). Amongst articles attempting some analysis of the strengths and weaknesses of the TRIF, critical views clearly formed the majority (Table 2). Care is needed in interpreting Table 2, because *status quo* bias (Porter and McIntyre 1984) may mean that those who find the TRIF useful don't bother to publish, while a minority of vocal critics appear as a majority. However, despite this possibility, Table 2 leads to the conclusion that there are serious limitations inherent in the TRIF, and that there are compelling reasons to overhaul or replace this indicator which is often assumed to be a valid proxy for journal quality.

**Table 2**. Summary of recent literature on the journal impact factor.

| Aspect | Weakness | Strength |
|---|---|---|
| Indication of Journal 'quality' | Unreliable correlation between TRIF and independent measures of evidence (Bain & Myles 2005) and quality (Seglen 1989, Woolgar 1991, Seglen 1997, Bath et al 1998, Frank 2003, Walter et al 2003, Bernstam et al 2006, Maier 2006, Ogden & Bartley 2008, Glynn et al 2010, Goldstein & Maier 2010, Schlumm 2010). Citation rates (and hence TRIF) can be influenced by factors other than scientific merit (Knothe 2006, Calver & Bradley 2010, Perneger 2010). Weak correlation between TRIF and article rejecton rates (Kurmis & Kurmis 2006). Not at core of principal component analysis (Bollen et al 2009, Glanzel 2009). Year-to-year variation in rank is about 20% (Altmann & Gorman 199). Lacks robustness against single outliers (Rousseau 2002, Metze 2010).Incomplete and inadequate measure (Coleman 2007) 'Dodgy evaluation criteria' (Lawrence 2007). 'Seriously debased' (Williams 2007). | Correlates with expert opinion (Drew & Karpf 1981, Saha et al 2003, Yue et al 2007), and with level of evidence (Lau & Samman 2007, Obremskey et al 2007, Sonderstrup-Andersen & Sonderstrup-Andersen 2008 Minelli et al 2009). Consistent with overall citation count (Hansen and Henriksen 1997, Callaham et al 2002, Chapman et al 2009, Haslam & Koval 2010, Hunt et al 2010) and with use of articles (Wulff & Nixon 2004). Overall, a reasonable measure of quality if used correctly (Schoonbaert & Roelants 1996, Lomnicki 2003, Cartwright & McGhee 2005, Gluud et al 2005, Racki 2009, Ruiz et al 2009, Abramo et al 2010). TRIF reflects likelihood of having a formal misconduct policy (Resnik et al 2009, Resnik et al 2010). |
| Rigour of TRIF | Open to manipulation (Rousseau & Van Hooydonk 1996, Garfield 1999, Jones 2002, Jones 2003, Kurmis 2003, Dong et al 2005, Monastersky 2005, Cheek et al 2006, Chew et al 2007, Smith 2007, Yu & Wang 2007, Campbell 2008, Falagas & Alexiou 2008, Mullen 2008, Ogden & Bartley 2008, Reedijk & Moed 2008, Archambault & Lariviere 2009, Brumback 2009a, Dempsey 2009, Hernan 2009, Ramsden 2009, Kapeller 2010, Krell 2010, Statzner & Resh 2010, Foo 2011, Mehrad & Goltaji 2011, Yu et al 2011). Different editorial policies influence TRIF (Moed et al 1996, Scully and Lodge 2005, Foo 2009). About 18% of journals have self-citation exceeding 20% (Ha et al 2006), and self-citation is significantly correlated with TRIF (Straub & Anderson 2009, Kurmis & Kurmis 2010, Mehrad & Goltaji 2010). Self-citation contributes fluctuation in TRIF (Leutner & Wirth 2007, Moller et al 2010, Campanario 2011b). TRIF lacks transparency (Van Driel et al 2008) | No proof of widespread manipulation (Andrade et al 2009). Editorial citations contribute little to TRIF (Campanario et al 2011). Raising quality of journal is best way to improve TRIF (Wang et al 2010). |
| Normalization | TRIF differs across disciplines and warrants normalization (Ugolini et al 1997, Coleman 1999, Pudovkin & Garfield 2004, Sombatsompop et al 2004, Sombatsompop & Markpin 2005, Rezaei-Ghaleh & Azizi 2007, Althouse et al 2009, Wagner 2009, Leydesdorff & Opthof 2010, Owlia et al 2011) | |
| Timeframe | Two years too short (van Leeuwen et al 1999, McGarty 2000, Sombatsompop et al 2004, Smith 2007). Variable-length may be needed to reflect disciplinary norms (Vanclay 2009). | 5-year impact factors may follow a similar (Campanario 2011a) or complementary pattern (Jacso 2009) to the 2-year TRIF. |
| Distribution and statistical assumptions | Non-normal distribution (Weale et al 2004). Journals cannot be ranked with great precision (Greenwood 2007). No statistics to inform signifance (Leydesdorff & Opthof 2010) | Possible to estimate standard errors (Schubert & Glanze 1983) |
| Database issues | Database problems may be a major source of bias (Tam et al 2006). Language bias may arise from the journals scanned (Kotiaho et al. 1999, Winkmann & Schweim 2000, Winkmann et al 2002, Taylor et al 2008, Schopfel & Prost 2009, Xiao et al 2009, Poomkottayil et al 2011), from insular citing patterns (Jacobs & Ip 2005, Stiftel & Mukhopadhyay 2007), or from errors arising from surname conventions (Meneghini et al 2008, Kumar et al 2009). Journals with similar titles may be incorrectly dealt in a single TRIF (Lange 2002). Some 25-35% of all citations contain errors (Todd et al 2010, Awrey 2011). "information available in the Science Citation Index is a rather unreliable indication" (Neuhaus 2009). | |
| Unintended consequences | Threatens viability of specialist journals (Zetterstrom 1999, Johnstone 2007) and disciplines (Brown 2007). May distort publication patterns away from prime audience (Postma 2007, Kapeller 2010). Intrinsic and deliberate errors are transferred to other decisions (Monastersky 2005, Starbuck 2005, Todd & Ladle 2008). May shift editorial focus to increase TRIF and away from other aspects of 'quality' (Ketcham 2008). | |

In summary, Table 2 reveals considerable concern that the TRIF is not a reliable indicator of journal quality, that it lacks rigour and requires normalizing before comparisons are attempted, that its 2-year timeframe is too short for meaningful trends to be established, that it lacks statistical validity, suffers database problems, and leads to problematic (if unintended) consequences. To put it bluntly, the TRIF "plays a particularly significant role in choosing a journal, and yet it is a controversial and, some would say, flawed metric" (Soreide & Winter 2010), an "outmoded surrogate for quality" (PLoS Medicine Editors 2006), "an ill-defined and manifestly unscientific number" (Rossner et al 2007) with "bias [that] originates mainly from misuse as well as abuse" (Falagas & Alexiou 2008). Others are more critical, concluding that the TRIF is "seriously debased … inescapable conclusion is therefore that the impact factor is worthless … should be killed off, and the sooner the better." (Williams 2007). "Scientists should be outraged that the worth of science is being measured by a secretive proprietary metric … should renounce the Thomson Reuters impact factor" (Brumback 2009b). Mode and Plume (2011) summarised more constructively, noting that the TRIF is a first generation bibliometric indicator, whilst the state-of-the-art is now third generation. It is time to progress to something better.

Apart from the limitations of the TRIF, a further problem is the behavioural change that it influences, along with unintended consequences. For instance, Lehmkuhl et al (2009) opined that "the goal is to increase the impact factor", but in reality, the goal should be to increase quality, and the TRIF should be a fortunate consequence of the better quality of articles. However, Seglen (1992) argued that that there was no evidence of a correlation between journal quality and the TRIF. Of concern is the increasing trend to the use of the TRIF for assessing the performance of individuals (Favaloro 2009), because the TRIF is also unsuited for this purpose (Epstein 2004).

It is apparent that many authors, even those involved with citation analysis, do not have an adequate understanding of these limitations (e.g. "These issues are rarely understood", Holsapple 2009). Thus it is appropriate to examine more closely some of the weaknesses of the TRIF.

## An evaluation of the Thomson Reuters Impact Factor (TRIF)

There are several aspects of the Garfield index that warrant closer attention. For convenience, these are grouped broadly as data errors, system faults, sampling deficiencies and statistical shortcomings. Collectively, these cast doubt on the ability to reproduce TRIFs as published in JCR with data from WoS or from comparable third-party data providers (Table 3). Attempts to reproduce the TRIF (as published in the JCR) with WoS data may differ two-fold (e.g., *American Journal of Bioethics*, Table 3). Depending on the nature of editorial material, the values obtained from different forms of the Garfield index may vary four-fold, even when compiled from the same database (*American Journal of Bioethics*, Table 3). Values for the Garfield index estimated from different databases may vary six-fold (*Australian Meteorological and Oceanographic Journal*, Table 3). Such observations cast doubt on the relevance of publishing TRIF with three decimal points.

**Table 3**. Comparison of variants of the Garfield index and of different databases

| Journal | Variants of Garfield index using Web of Science | | | | TRIF from JCR | $G_{am}$ from alternative databases | |
|---|---|---|---|---|---|---|---|
| | $G_{mm}$ | $G_{am}$ | $G_{11}$ | $G_{11}$ (no self) | JCR | Scopus | Scholar |
| *Rev Mod Phys* | 46.629 | 50.610 | 43.622 | 43.402 | 51.695 | 47.506 | † |
| *Nature* | 11.744 | 32.573 | ‡ | ‡ | 36.101 | 33.079 | † |
| *Am J Bioethics* | 0.415 | 1.630 | 1.959 | 1.932 | 3.986 | 0.707 | 2.590 |
| *World J Gastroenterology* | 1.923 | 1.835 | ‡ | ‡ | 2.240 | 2.406 | † |
| *Forestry* | 1.205 | 1.218 | 1.310 | 1.149 | 1.460 | 1.725 | 1.568 |
| *Aust Forestry* | 0.818 | 0.818 | 0.500 | 0.471 | 0.836 | 0.786 | 0.400 |
| *Aust Meteor & Oceanogr J* | 0.662 | 0.593 | 0.525 | 0.390 | 0.179 | 0.473 | 2.949 |

‡ WoS limits output of cited references to 500 records. † Google Scholar limits output to 1000 records.

This study was triggered through curiosity surrounding the discovery of an error in the handling of citations to a 2008 paper in *Forestry*, and an exploration of how that error might affect the 2010 TRIF. That exploration revealed more errors, which in turn led to an analysis of several journals spanning extremes (high and low TRIF, many and few citations, etc.), *all* of which revealed errors in the WoS data. Undoubtedly, sampling these extremes increased the likelihood of revealing errors, so this study should be seen as indicative rather than representative of all data within the WoS, but it does raise several matters of concern, and aligns with calls for an independent evaluation of JCR estimates of impact (Rossner et al 2007).

*Gross Errors*

At the time of writing (August 2011), WoS contained two records of a paper in *Forestry* by Skovsgaard and Vanclay (2008), both incorrect (Figure 6; correct pagination is 13-31). Curiously, the second, more incomplete record included the correct digital object identifier (DOI), and recognised 20 citing documents most of which contained the correct details, so Figure 6 reflects data entry errors by WoS, not typographic errors by citing authors. These errors were drawn to the attention of WoS during the preparation of this manuscript, and these errors were quickly amended. A comparable example is also evident with an earlier article in the journal *Science* (Figure 6), and many other examples of such errors can be found. Figure 6 illustrates errors due largely to WoS encoding, but author errors may also contribute substantially to incorrect references (Simkin and Roychowdhury 2003).

| Cited Author | Cited Work [SHOW EXPANDED TITLES] | Year | Volume | Page | Article ID | Citing Articles ** | View Record |
|---|---|---|---|---|---|---|---|
| Skovsgaard, J. P. | FORESTRY | 2008 | 81 | 12 | | 2 | View Record |
| SKOVSGAARD JP | FORESTRY | 2008 | 81 | 13 | 10.1093/forestry/cpm041 | 20 | |

| Cited Author | Cited Work [SHOW EXPANDED TITLES] | Year | Volume | Page | Article ID | Citing Articles ** | View Record |
|---|---|---|---|---|---|---|---|
| Vanclay, JK | SCIENCE | 2001 | 293 | 1007 | | 17 | View Record |
| VANCLAY JK | SCIENCE | 2001 | 293 | NIL1 | | 2 | |
| VANCLAY JR | SCIENCE | 2001 | 293 | A1007 | 10.1126/science.293.5532.1007a | 2 | |

**Figure 6**. Illustration of some errors in WoS records of *Forestry* articles.

The error rate amongst journal articles is worrisome enough, but error rate escalates for citations with books and conference proceedings. Figure 7 illustrates a diverse range of errors introduced into the title of a single book, *Modelling Forest Growth and Yield*. Some of these errors are faithful reproductions of errors made by citing authors, but many appear to be data-entry errors introduced by Thomson Reuters (or its predecessors). Figures 6 and 7 feature problems with citations to the author's work, but the issues illustrated are common and similar issues can be reproduced for many other authors.

| Cited Author | Cited Work [SHOW EXPANDED TITLES] | Year | Volume | Page | Article ID | Citing Articles ** | View Record |
|---|---|---|---|---|---|---|---|
| VANCLAY JK | MODELING FOREST GROW | 1994 | | | | 22 | |
| VANCLAY JK | MODELING FOREST YIEL | 1994 | | | | 1 | |
| VANCLAY JK | MODELLIGN FOREST GRO | 1999 | | | | 1 | |
| VANCLAY JK | MODELLIN GFOREST GRO | 1994 | | | | 1 | |
| VANCLAY JK | MODELLING FOREST GRO | 2001 | | | | 1 | |
| VANCLAY JK | MODELLING FOREST GRO | 1999 | | | | 2 | |

**Figure 7**. Selected WoS errors in the title of a book, *Modelling Forest Growth and Yield* published in 1994.

The errors illustrated in Figures 6 and 7 are minor typographic errors, but other errors exist in WoS. For instance, amongst the citations contributing to the 2010 TRIF for *Australian Forestry* (Table 1) is an incorrect citation from *European Journal of Pharmaceutical Sciences*, in which a paper by M Thommes and others was apparently confused with a different *Australian Forestry* paper by D Thomas (Figure 8). It is unclear how this error may have arisen, but it appears that the wrong author was selected from a drop-down list during data-entry.

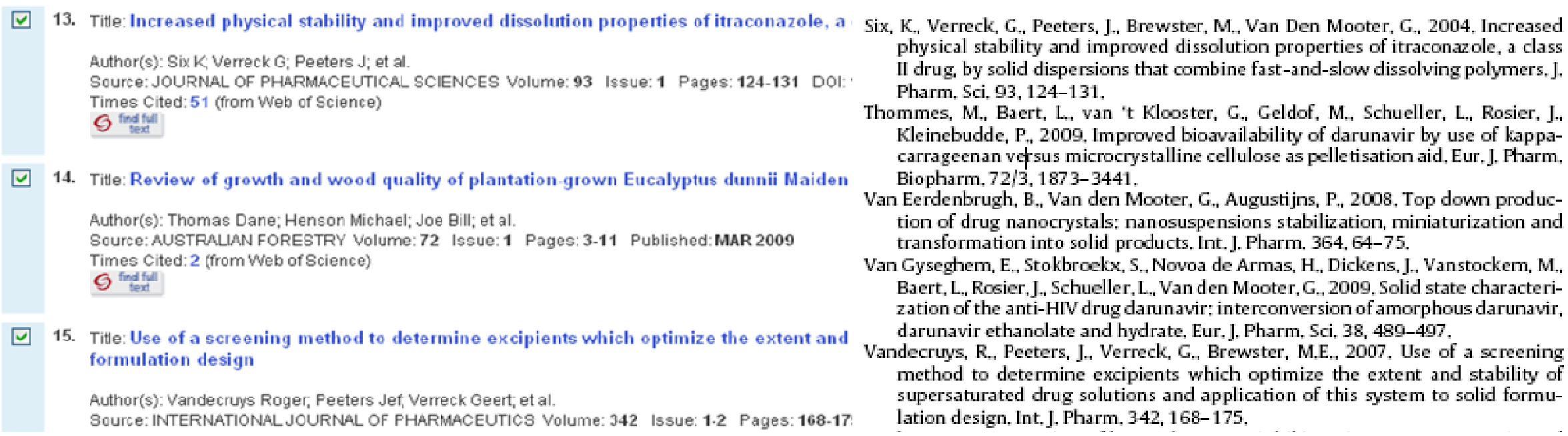

13. Title: Increased physical stability and improved dissolution properties of itraconazole, a
Author(s): Six K; Verreck G; Peeters J; et al.
Source: JOURNAL OF PHARMACEUTICAL SCIENCES Volume: 93 Issue: 1 Pages: 124-131 DOI:
Times Cited: 51 (from Web of Science)

14. Title: Review of growth and wood quality of plantation-grown Eucalyptus dunnii Maiden
Author(s): Thomas Dane; Henson Michael; Joe Bill; et al.
Source: AUSTRALIAN FORESTRY Volume: 72 Issue: 1 Pages: 3-11 Published: MAR 2009
Times Cited: 2 (from Web of Science)

15. Title: Use of a screening method to determine excipients which optimize the extent and formulation design
Author(s): Vandecruys Roger; Peeters Jef; Verreck Geert; et al.
Source: INTERNATIONAL JOURNAL OF PHARMACEUTICS Volume: 342 Issue: 1-2 Pages: 168-17

Six, K., Verreck, G., Peeters, J., Brewster, M., Van Den Mooter, G., 2004. Increased physical stability and improved dissolution properties of itraconazole, a class II drug, by solid dispersions that combine fast-and-slow dissolving polymers. J. Pharm. Sci. 93, 124–131.

Thommes, M., Baert, L., van 't Klooster, G., Geldof, M., Schueller, L., Rosier, J., Kleinebudde, P., 2009. Improved bioavailability of darunavir by use of kappa-carrageenan versus microcrystalline cellulose as pelletisation aid. Eur. J. Pharm. Biopharm. 72/3, 1873–3441.

Van Eerdenbrugh, B., Van den Mooter, G., Augustijns, P., 2008. Top down production of drug nanocrystals: nanosuspensions stabilization, miniaturization and transformation into solid products. Int. J. Pharm. 364, 64–75.

Van Gyseghem, E., Stokbroekx, S., Novoa de Armas, H., Dickens, J., Vanstockem, M., Baert, L., Rosier, J., Schueller, L., Van den Mooter, G., 2009. Solid state characterization of the anti-HIV drug darunavir: interconversion of amorphous darunavir, darunavir ethanolate and hydrate. Eur. J. Pharm. Sci. 38, 489–497.

Vandecruys, R., Peeters, J., Verreck, G., Brewster, M.E., 2007. Use of a screening method to determine excipients which optimize the extent and stability of supersaturated drug solutions and application of this system to solid formulation design. Int. J. Pharm. 342, 168–175.

**Figure 8.** Reference list in WoS (left) and original (right) of Van Gyseghem (2010).

Yet another error arises from the large number of test records that remain littered throughout the WoS database. Figure 9 illustrates one such case, where reference 1 is clearly a partial reconstruction of reference 2 used for testing purposes, but which was not removed at the end of testing. There is potential for these remnants of testing to inflate the TRIF for the journal *Test* (ISSN 1133-0686), but the 2010 TRIF appears unaffected.

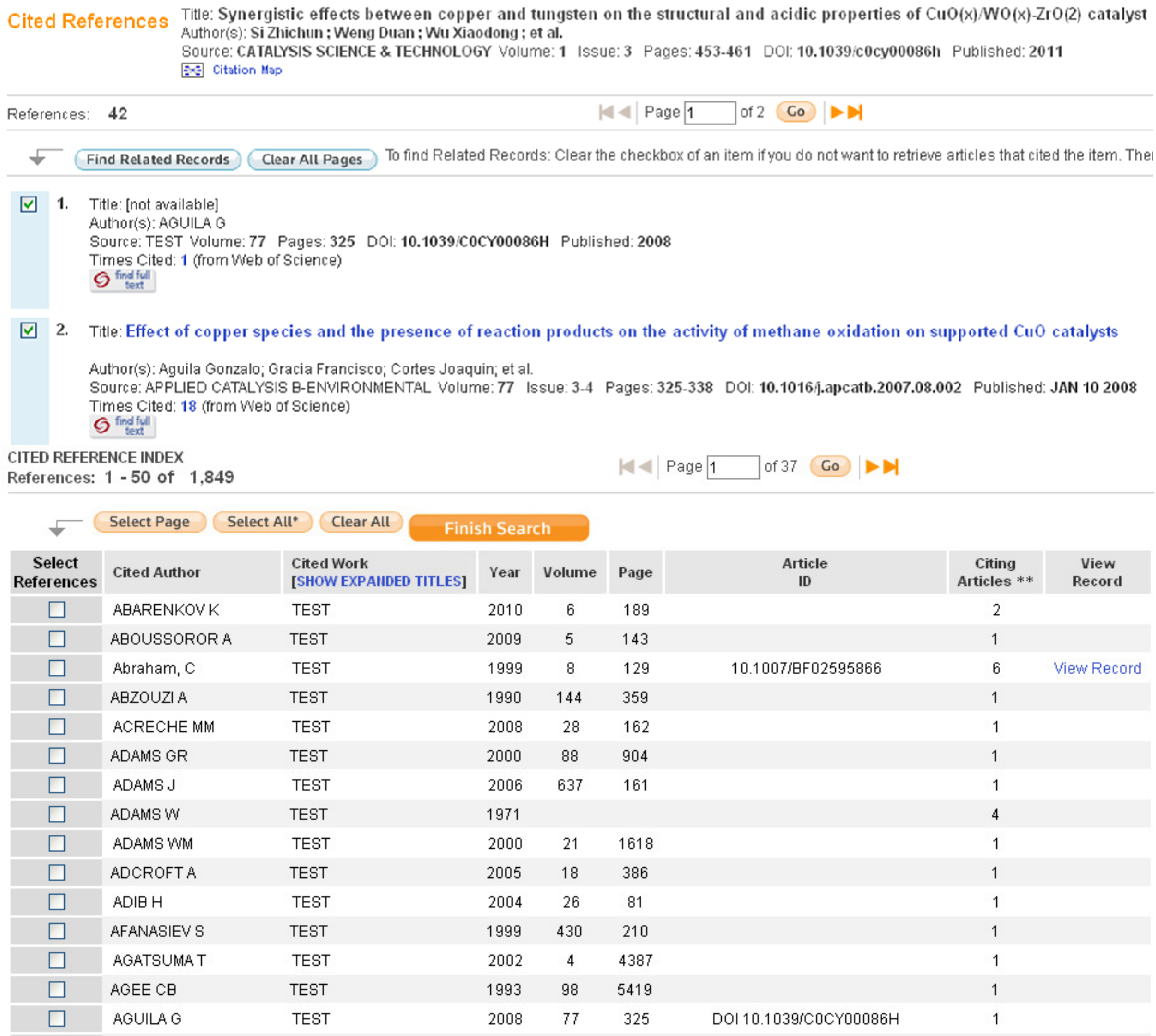

Cited References

Title: Synergistic effects between copper and tungsten on the structural and acidic properties of CuO(x)/WO(x)-ZrO(2) catalyst
Author(s): Si Zhichun ; Weng Duan ; Wu Xiaodong ; et al.
Source: CATALYSIS SCIENCE & TECHNOLOGY Volume: 1 Issue: 3 Pages: 453-461 DOI: 10.1039/c0cy00086h Published: 2011
Citation Map

References: 42 — Page 1 of 2 Go

Find Related Records | Clear All Pages — To find Related Records: Clear the checkbox of an item if you do not want to retrieve articles that cited the item. The

1. Title: [not available]
Author(s): AGUILA G
Source: TEST Volume: 77 Pages: 325 DOI: 10.1039/C0CY00086H Published: 2008
Times Cited: 1 (from Web of Science)

2. Title: Effect of copper species and the presence of reaction products on the activity of methane oxidation on supported CuO catalysts
Author(s): Aguila Gonzalo; Gracia Francisco; Cortes Joaquin; et al.
Source: APPLIED CATALYSIS B-ENVIRONMENTAL Volume: 77 Issue: 3-4 Pages: 325-338 DOI: 10.1016/j.apcatb.2007.08.002 Published: JAN 10 2008
Times Cited: 18 (from Web of Science)

CITED REFERENCE INDEX
References: 1 - 50 of 1,849 — Page 1 of 37 Go

Select Page | Select All* | Clear All | Finish Search

| Select References | Cited Author | Cited Work [SHOW EXPANDED TITLES] | Year | Volume | Page | Article ID | Citing Articles ** | View Record |
|---|---|---|---|---|---|---|---|---|
| ☐ | ABARENKOV K | TEST | 2010 | 6 | 189 | | 2 | |
| ☐ | ABOUSSOROR A | TEST | 2009 | 5 | 143 | | 1 | |
| ☐ | Abraham, C | TEST | 1999 | 8 | 129 | 10.1007/BF02595866 | 6 | View Record |
| ☐ | ABZOUZI A | TEST | 1990 | 144 | 359 | | 1 | |
| ☐ | ACRECHE MM | TEST | 2008 | 28 | 162 | | 1 | |
| ☐ | ADAMS GR | TEST | 2000 | 88 | 904 | | 1 | |
| ☐ | ADAMS J | TEST | 2006 | 637 | 161 | | 1 | |
| ☐ | ADAMS W | TEST | 1971 | | | | 4 | |
| ☐ | ADAMS WM | TEST | 2000 | 21 | 1618 | | 1 | |
| ☐ | ADCROFT A | TEST | 2005 | 18 | 386 | | 1 | |
| ☐ | ADIB H | TEST | 2004 | 26 | 81 | | 1 | |
| ☐ | AFANASIEV S | TEST | 1999 | 430 | 210 | | 1 | |
| ☐ | AGATSUMA T | TEST | 2002 | 4 | 4387 | | 1 | |
| ☐ | AGEE CB | TEST | 1993 | 98 | 5419 | | 1 | |
| ☐ | AGUILA G | TEST | 2008 | 77 | 325 | DOI 10.1039/C0CY00086H | 1 | |

**Figure 9**. Artifacts from testing remaining in the WoS database. Top: correct and test variants of citation to work by Aguila. Bottom: Only one correct record amongst first 15 records in a search for citations to the journal 'Test' which returned 1849 citations.

Thus it is useful to identify four kinds of problematic citation. Citations found in a citing document may be

1. correct and complete (in this case, WoS creates a 'ViewRecord' link, e.g., Figures 6 and 7),
2. partially correct but incomplete so that an unambiguous link cannot be made without further research (typically minor typographic errors, e.g., Figure 7),
3. faulty (incorrect but complete) in such a way that a reference has been erroneously linked to the wrong cited document (potentially, there are two forms: the difficult-to-detect situation where an author incorrectly attributes a fact to the wrong document, and the example in Figure 8 where a problem in WoS has 'crossed wires' and wrongly linked a valid reference with a different source document), and
4. 'ghostly' in the sense that a reference is made to a document not seen by WoS, either because the cited document is not in a journal scanned by WoS, or because the document does not exist (incorrect and incomplete).

These four possibilities embrace the four combinations of (in)complete and (in)correct links.

**Table 4.** Citations during 2010 to *Forestry* articles published in 2008-09

| Category | Correct | Incorrect | Total |
|---|---|---|---|
| Complete | 113 | 2 | 115 |
| Incomplete | 5 + 7 | 1 | 13 |
| Total | 125 | 3 | 128 |

It is difficult to establish which of these four possibilities the TRIF relies upon. For instance, Table 4 illustrates a futile attempt to reconstruct the TRIF for the journal *Forestry*. The JCR reveals that during 2010, 127 citations were made to 87 papers published in 2008-09, indicating a TRIF of 1.460. However, a search of WoS using the cited reference search reveals 143 citations, including 125 correct citations and 15 problematic citations (Table 4), suggesting an impact factor of 1.3 to 1.5, depending on assumptions. It is instructive to examine the source of errors reported in Table 4. Seven of the incomplete citations arose because of a single-digit typographic error in the citing paper, with WoS faithfully reproducing this error (*Forestry* is not unique in this regard; Figure 10 illustrates a corresponding example from *Nature*). One of the incorrect citations arose through an author error: Tal and Gordon (2010) referred to a 'ghost' paper, 'The influence of early planting on the sapling success rate and the development' in *Forestry* 10:35–37, which actually appeared as 'Influence of Autumn Planting on Forest Tree-Seedlings Survival and Growth' in *Forest: Journal of Forests, Woodlands & Environment* 10:35-37 (2008), not seen by WoS. Two of the incorrect citations arise because of miscoding by WoS: Medarević et al (2010) cited a paper as "*Forestry* 3, 17-31 (in Serbian)" that was in reality *Šumarstvo* 60(3):17-30 (ISSN 0350-1752) and was encoded by WoS as *Forestry* 3:17 (2008) despite a complete mismatch of year and volume number for the journal *Forestry* (ISSN 0015-752X). And Preti et al (2010) cited a paper correctly as "*iForest: Biogeosci. Forestry* 1 (2008) 141–144" that was encoded by WoS as *Forestry* 1:141 (2008), again despite the mismatch between year and volume number. Clearly, WoS suffers from some substantial coding errors. Because of these errors, and because of updates to the WoS database since the JCR was released, it is impossible to independently reproduce the TRIF (for *Forestry*, or for any journal reported in Table 3), or to empirically establish whether any erroneous citations are omitted from the TRIF calculation.

| | | | | | | | |
|---|---|---|---|---|---|---|---|
| ABDO | NATURE | 2009 | 462 | 331 | | 1 | |
| Abdo, A. A. | NATURE | 2009 | 462 | 331 | 10.1038/nature08574 | 73 | View Record |
| ABDO AA | NATURE | 2009 | 463 | 331 | | 1 | |
| ABDO AA | NATURE | 2009 | 426 | 331 | | 1 | |
| ABDO AA | NATURE | 2009 | | | | 1 | |
| ABDO MA | NATURE | 2009 | 462 | 331 | | 1 | |

**Figure 10**. Illustration of some errors in WoS records of *Nature* articles.

Typographic errors such as those illustrated in Figure 10 are common, and may be inevitable given that good researchers are not always good typists. But the frequency of such errors appears to vary considerably between journals, a strange phenomenon since the cited journal has little control over the attention to detail in the citing journal (except in the special case of self-citation). One journal with a remarkably low incidence of such errors is *Molecular Ecology*, previously mentioned because of its self-citation practices. WoS reveals remarkably few instances of such errors that might affect their TRIF. One instance occurred in 2008 when the *Journal of Systematics and Evolution* erroneously referred to "*Molecular Ecology* 15:897–915" (2006) instead of *Molecular Ecology* 15(13):4065–4083 (2006). Another occurred in a 2009 author self-citation when *Zoologica Scripta* referred to "*Molecular Ecology*, online" (2008) instead of to *Molecular Ecology* 17(24):5205–5219 (2008). Other search engines reveal author errors in citations to *Molecular Ecology*, but since these are not seen by WoS, these errors do not affect the TRIF. For instance, Robert (2010) mistakenly used 2008 in conjunction with *Molecular Ecology* 18(2), 319–331 (2009), an understandable mistake since the article was "first published online" in 2008. Two papers currently 'online first' in *Biological Invasions* incorrectly cite "Invasion genetics of the round goby: tracing Eurasian source populations to the New world" instead of the correct "Invasion genetics of the Eurasian round goby in North America: tracing sources and spread patterns". And three articles, including an author self-citation in *Molecular Ecology* use the title "Landscape genetic structure of tailed frogs in protected versus managed forests" instead of the correct "Landscape genetic structure of coastal tailed frogs (*Ascaphus truei*) in protected vs. managed forests" – but in the latter two illustrations, since year, volume and page numbers are correct, they would not affect the TRIF of *Molecular Ecology*. There is a slender possibility that authors interested in molecular ecology are more fastidious regarding the detail of citations than the general population of scientists, but this would not explain the absence of WoS-introduced errors. This low error rate for citations to *Molecular Ecology* suggests that its editors work closely with WoS to identify and correct any errors, a good strategy for TRIF-conscious editors.

In an ideal world, an impact factor would rely only on complete and correct citations, reinforcing quality control through the whole journal publication chain. However, it appears that the 2010 JCR merely tallies up for each journal, all the citations made to papers published in 2008-09, irrespective of whether these are complete, correct, or ghost. This offers an entirely new possibility to manipulate the TRIF: as well as citing many recent articles (e.g., Rieseberg et al 2011), editors could cite ghost articles that could usefully increase a journal's TRIF without distorting the performance indicators for real contributors. It also raises the prospect of another measure of journal quality control: could the frequency of citation errors in the citing journal be useful as a proxy for journal quality? Such a proxy would not only reflect on the rigour of editorial procedures (in checking errors), but also in 'after sales service' of the collaboration between a journal and WoS (or other provider) to correct subsequent incoming citations.

Given the lax error-checking by WoS, it is tempting to include a series of ghost articles in a review of this kind to demonstrate weaknesses of the TRIF (but this has not been done in the present case). Many journals receive relatively few citations (e.g., *Australian Forestry* received about 46 citations in 2010, Table 1, and *Mediterranean Politics* received fewer than 20 citations in 2010), so it can be a relatively easy matter to double the TRIF with a single review or editorial with a series of references to ghost papers that would not fool a genuine researcher (e.g., possibly concerning *Eucoliptus sokal*, a non-existent and deliberately misspelled plant species commemorating a hoax by Alan Sokal). A series of such ploys by sceptical editors could completely undermine the TRIF in its current form and force action on Thomson-Reuter's part (possibly suspension of the citing journal, but hopefully a revision of the TRIF to the $G_{11}$ basis).

*System Faults*

In addition to the ad hoc data entry errors identified above, the WoS and JCR suffer several system errors. Perhaps the most conspicuous of these is the handling of dates when journals change in some way (e.g., new addition, change of name, re-commenced after suspension).

For instance, one of the more conspicuous errors arises with *PLOS One*. *PLOS One* was first published in 2006, but WoS records 19 citations to *PLOS One* articles published in 2005 that are presumably erroneous (Figure 11). One of these citations includes a DOI that indicates a 2010 article; others have unique article numbers (shown as page numbers) that can resolve the anomalous year of publication.

| Cited Author | Cited Work [SHOW EXPANDED TITLES] | Year | Volume | Page | Article ID | Citing Articles ** | View Record |
|---|---|---|---|---|---|---|---|
| AHO T | PLOS ONE | 2005 | 5 | | | 1 | |
| ALEXIOU P | PLOS ONE | 2005 | 5 | E9171 | | 1 | |
| AYENEW A | PLOS ONE | 2005 | 5 | E9702 | | 1 | |
| BALCAN D | PLOS ONE | 2005 | 2 | E501 | | 1 | |
| BANSAL S | PLOS ONE | 2005 | 5 | E9360 | | 1 | |
| COLL M | PLOS ONE | 2005 | 5 | | DOI 10.1371/JOURNAL.PONE.0011842 | 1 | |
| DELAFUENTE IM | PLOS ONE | 2005 | 5 | | | 1 | |
| DERE R | PLOS ONE | 2005 | 5 | E9239 | | 1 | |
| GOLEMBA MD | PLOS ONE | 2005 | 5 | E8751 | | 1 | |
| GOWEN B | PLOS ONE | 2005 | 3 | E3725 | | 1 | |
| KANO S | PLOS ONE | 2005 | 5 | | | 1 | |
| LU R | PLOS ONE | 2005 | 5 | | | 1 | |

**Figure 11**. Selected mis-matched citations from WoS referring to *PLoS One* in the year before publication commenced.

Table 2 reveals a considerable difference in the various Garfield indices for the *Australian Meteorological and Oceanographic Journal*. These differences are due in part to the re-naming of the journal from the *Australian Meteorological Magazine*. WoS reports TRIF for both (0.179 and 0.935 respectively), but is seems more appropriate to combine these (0.576) as in all other respects it remains the same journal. Change of a journal name is a relatively common event with about 30 such changes each year, so the TRIF should accommodate them efficiently.

The *World Journal of Gastroenterology* was suspended from WoS during 2005-07, apparently due to allegations of self-citation, and was reinstated in 2008. The JCR calculation of the 5-year impact factor includes 2313 cites to articles published in 2005, but TRIF recognises no documents from 2005, so the denominator is underestimated and the TRIF is biased upwards. The TRIF calculation should compare like with like, either by including 2005 documents or omitting 2005 citations. In any case, the practice of gratuitous editorial self-citation continues in other journals: for instance, *Molecular Ecology* recently published an editorial with 179 self-citations (Rieseberg et al 2011). Science would be better served if the TRIF omitted self-citations (either to the offending journal, or generally to all journals) rather than suspending a few of the offenders.

*Sampling Deficiencies*

The fact that WoS is a sample of scientific literature is often overlooked, and the TRIF is often treated as if it was based on a census. In reality, WoS draws on a sample of the scientific literature, selected following their own criteria (Anon 2011), as amended from time to time (e.g., through suspensions for self-citation). Other providers (Scopus, Google Scholar) and evaluation agencies (e.g., the Excellence for Research in Australia, ARC 2010) utilize different samples of the scientific literature, so their interpretation of a corresponding impact factor would differ from the TRIF. And WoS policies to include or suspend a journal also affect the TRIF. Because *World Journal of Gastroenterology* was suspended in 2005, WoS has no data, but Scopus indicates that this journal made over 6000 citations to articles from 2004-05, with the result that the suspension of one journal could have deflated the TRIF for other gastroenterology journals by as much as 1%. These sources of variation lead one to question the practice of publishing the TRIF with three decimal points, and to ask why there is no statement regarding variability.

The TRIF also represents a temporal sample, evaluating the citations received in one calendar year, by work published in the two previous calendar years. For some disciplines, this 2-year window is appropriate, but for others it is inadequate (Seglen 1997, Vanclay 2009, Bensman et al 2010). Figure 12 illustrates that patterns of citation accrual can vary greatly: the 1998 TRIF sampled the mode of citations to 1996 articles in *Nature*, but the mode is not reached for *Ecology* until a decade has elapsed (Figure 12). The 1998 TRIF would have secured a reliable indication of *Nature*'s impact, but only the tip of the iceberg in *Ecology*, greatly underestimating its impact. The release of a 5-year impact factor only partly alleviates this problem. Furthermore, the disconnect between the 2-year window of the TRIF and the decade taken for some journals to peak (e.g., *Ecology*, Figure 12) introduces yet another way to manipulate the TRIF by allowing contributions to appear informally on-line before releasing the official date-stamped print version. Such a practice offers little advantage for a journal that peaks early (e.g., *Nature*, Figure 12), but may substantially alter the TRIF of slow-to-peak journals such as *Ecology*. At the time of writing, *Ecology* displayed a 4-month queue of 66 articles displayed on-line ahead of print, a tactic probably designed to facilitate rapid science communication, but which has the convenient benefit of inflating the TRIF.

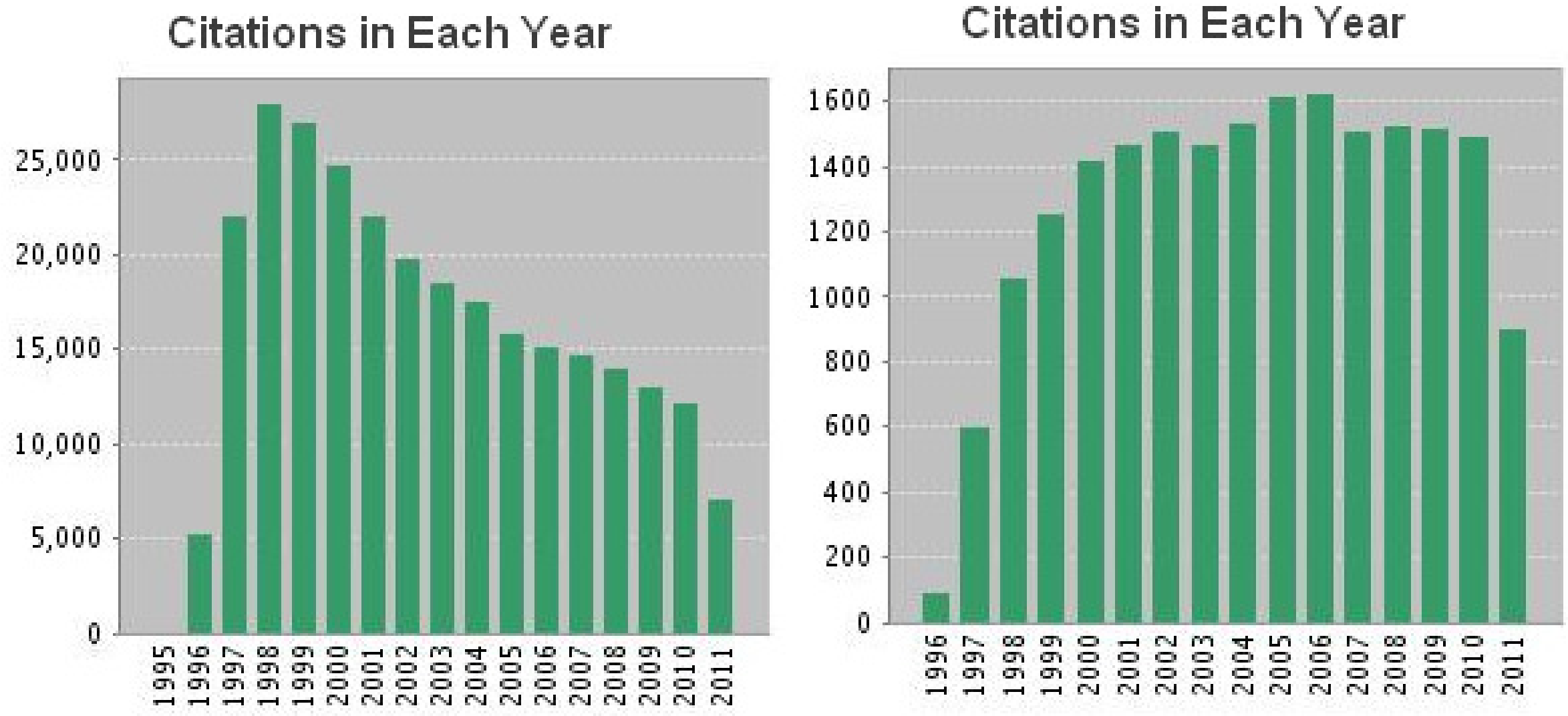

**Figure 12**. Aggregate citations to articles published in 1996 in *Nature* (left) and *Ecology* (right), illustrating different trends in citation accrual (Images from Web of Science).

*Statistical Shortcomings*

The usual distribution of citations is highly skewed, with a few articles often-cited and many articles rarely-cited. The utility of the mean, a conventional statistic with normal distributions, does not apply to such distributions. With a normal distribution (such as would be expected with e.g., adult body mass), the mode, mean and median all have similar values. However, with citation data, these common four parameters may differ dramatically (Table 5). By way of analogy, the pattern of citations is more like that of scratch lotto tickets than a normal distribution. In this analogy, each article is a ticket, and the citations are the payout. The mode should be of interest to a ticket-buyer (cf. the reader or impact assessor), because it reflects most likely prize. But most players overlook the mode, and focus on the major prizes, even though the chance associated with these prizes is small. The mean may be of interest to the auditor (cf. librarian), because it reflects the total payout if all the tickets are sold. The median is of little utility, except to an advertiser who can claim that half the prizes exceed a certain value. Continuing with the analogy, the *Gmm* takes into account that some ticket-sellers add extra incentives to try to increase their sales (as do editors, with strategic self-citations). The implication is that many users of impact statistics may be better advised to use the mode (the most likely citation count) rather than the Garfield index which is a biased estimate, ill-suited to a distribution of this nature. There is a long history of statistical misuse in science (Cohen 1938), but citation metrics should not perpetuate this failing.

**Table 5**. Selected parameters for citations accruing to articles published 2008-09.

| Journal | *Gmm* | Mean | Median | Mode | No. of articles & reviews |
|---|---|---|---|---|---|
| *CA-Cancer* | 209.5 | 122.2 | 6 | 0 | 72 |
| *PLOS Pathogens* | 17.9 | 17.3 | 13 | 9 | 744 |
| *Mol Ecol* | 14.8 | 13.2 | 9 | 5 | 868 |
| *Plant Phys* | 14.4 | 13.5 | 11 | 6 | 1073 |
| *J Hydro* | 5.5 | 5.4 | 4 | 3 | 923 |
| *Nat Prod Comm* | 1.8 | 1.8 | 1 | 0 | 691 |

## Way forward

This review of literature and empirical evidence reveals broad recognition that the TRIF is not a panacea, and leaves considerable room for improvement. It is important that the TRIF is improved, because it is influential in shaping science and publication patterns (Knothe 2006, Larivière & Gingras 2010). The advent of several alternative metrics (e.g., Eigenfactor, Article Influence Score, h-index) and providers (e.g., Scopus, SCImago) are both a welcome addition allowing users to choose the metric most suited to their needs, and a force for change, threatening the dominance of the single multi-purpose factor provided by Thomson Reuters. However, there remains a need for many of the 'gate-keeping' services that Thomson Reuters provides, in assessing timeliness of publication and the rigour of the review process. This creates the opportunity for Thomson Reuters (or new providers) to reposition such services in a way that is more constructive and supportive of good science.

The Garfield index had its origins in the desire to inform library subscription decisions (Garfield 2006), but it has gradually evolved into a status symbol for journals, which at its best is used to attract good manuscripts, and at worst is widely manipulated. It often serves as a proxy for journal 'quality', but is increasingly used more dubiously as a proxy for article quality (Postma 2007). It lacks transparency, repeatability and rigour. But despite all these failings, there remains a general perception that the TRIF is useful, probably because advocates have no better indicator of journal quality, an ill-defined notion about the value-added offered by a journal that is not reflected in other citation metrics. So what is the value-adding that a journal can offer an author's contribution in its journey from manuscript to published article? There is no doubt that value is added – most readers prefer a publisher's reprint over an author's preprint – but the elements of this value-adding are rarely defined. Elsevier (2011) defines this as peer-reviewing and "any other value added to it by a publisher (such as formatting, copy editing, technical enhancement and the like)", but this definition lacks a clear explanation of the elements of peer review. In common with Laband (1990, "value-adding by editors appears to derive principally from efficient matching of papers with reviewers"), this neglects the editorial role of checking for duplication, 'salami' (Abraham 2000), plagiarism and fraud. It is rarely made clear whether this checking is expected of the reviewers, completed by the editorial office, or adequately dealt with at all – and therein lies an unfilled need for scientific publication, and an opportunity for commercial providers. Science would be well served by an independent system to certify that editorial processes were prompt, efficient and thorough.

Roosendaal & Geurts (1997) argued that science communication involves four competing elements: registration, certification, awareness and archiving, which together form a value chain (Van de Sompel et al 2004). In many cases, the weakest link of this value chain is the certification that establishes that a work is a valid scientific contribution. There are several aspects involved (Table 6), but few of these are an integral part of the review process (Weller 2001, Hames 2007). Whilst some of these aspects are assured by self-interest and amenable to self-regulation, other important roles may be neglected, and there are several accounts illustrating the reluctance of editors to take decisive action on problematic articles (Lock 1995, Anon 2004, Chalmers 2006, Gollogly and Momen 2006), and Horner and Minifie (2011) draw attention to post-retraction citations that indicate deficiencies with the retraction process and other associated deficiencies. Many of the 'rigour and ethics' responsibilities (Table 6) are passed on to voluntary referees, who often lack the time and inclination to rigorously check for fraud and duplicate or 'salami' publication (Dost 200). Indeed, Bornmann et al (2008) point out that guidelines for referees rarely mention such aspects, and Wager et al (2009) observed that many science editors remain unconcerned about publication ethics and misconduct, despite the efforts of the Committee on Publication Ethics (Godlee 2004). Worse is the observation by Fox (1994) that “when reviewers do detect suspicious findings, they have been reluctant to alert editors”. Some editors seek to push ethical responsibilities back to the author (e.g., Abraham 2000, Tobin 2002, Roberts 2009), despite the prevalence of duplicate and fraudulent publications indicating that self-regulation by authors is insufficient (Gwilym et al 2004, Johnson 2006, Berquist 2008). There is little excuse for this complacency, as tools exist to assist the detection of plagiarism and fraudulence (e.g., McKeever 2006, Meyer zu Essen & Stein 2006, Rossner 2006,), and the consequences of some science fraud (e.g. medical ghost writing, e.g., Eaton 2005, Gøtzsche et al 2009, Barbour 2010) may be serious and far-reaching.

**Table 6**. Participant expectations regarding journal quality

| Viewpoint | Author | Reader | Service providers |
|---|---|---|---|
| Rigour & Ethics | Constructive feedback from editors and reviewers, on language, statistics and graphics. | Assurance of rigorous science, free of duplication, ‘salami’, plagiarism and fraud. | |
| Efficiency | Prompt editorial processes for efficient progression from submission to publication. | On-line access with click-through links to sibling, cited and citing articles. | On-line presence; Inclusion in bibliometric databases; Longevity |

There is a potential role for Google Scholar in helping to reduce fraud and plagiarism in science. Google Scholar already routinely displays “*n* versions of this article” in search results, and it could usefully display “other articles with similar text” and “other articles with similar images”. Such an addition would be very useful for researchers when compiling reviews and meta-analyses.

Clearly, good science requires a more proactive role from editorial offices, and the pursuit of this role is not reflected in the TRIF. In its present form, the TRIF is not fit for purpose, and does not serve the role of a journal ‘quality’ indicator often assumed by users. In broad terms, there are three options for the future of the impact factor:

1. The TRIF could be retained in a similar form, but amended to deal with its greatest limitations. It should move from $G_{mm}$ to $G_{11}$ (verified citations only) to reduce errors and to maintain pressure on the value chain for quality control, and should rely only on citations from articles and reviews, to articles and reviews. It should cease with 3-decimal detail, and should round factors appropriately according to the error associated with the estimate. The discontinuity introduced by these changes make it opportune to re-examine the timeframe, and to abandon the 2-year window in favour of an alternative that reflects the varying patterns of citation accrual in different disciplines.
2. Failing appropriate action by Thomson Reuters, the scientific community could rely on a community-based rating of journals, in much the same was as *PLoS One* does for individual articles, and as other on-line service providers offer to clients (e.g., TripAdvisor, Jeacle & Carter 2011).
3. Finally, the preferable option is for Thomson Reuters to abandon the TRIF in its present form, and to expand on other existing services that are currently implicit in the gate-keeping implicit in the inclusion in WoS. The scientific community would benefit from independent certification of journals, assuring not only timeliness and rigorous review (currently part of the gateway to be admitted to TRIF), but also to certify more broadly the standard of quality control (e.g., systematic checks for plagiarism, duplicate publication and fraud).

This paper is not the first to highlight the need for independent certification. In their review, Saunders & Savulescu (2008) called for “independent monitoring and validation” of research. There have been several calls (Errami & Garner 2008, Errami et al 2008, Butakov and Scherbinin 2009, Habibzadeh and Winker 2009, Foo 2010) for greater investment in, and more systematic efforts directed at detecting plagiarism and duplication. Callaham and McCulloch (2011) concluded that the monitoring of reviewer quality is even more crucial to maintain the mission of scientific journals. But despite these many calls for reform, change has been imperceptible. The TRIF remains essentially unchanged, but supplemented with a 5-year variant, and the Eigenfactor and Article Influence Score. And journals continue to tout their achievements when their TRIF increases, despite the dubious nature of the indicator.

The time has come to abandon the TRIF, and to replace it with a system that is better aligned with quality considerations in scientific publication. Such a system could be a rating system that allocated for instance one to to five stars for editorial efficiency and value-adding, for the rigour and constructiveness of the review process, and for procedures to detect and deal with plagiarism and other lapses of ethics. Such a system could be a powerful force for improving science: stars should not be awarded for a high manuscript rejection rate, but for the extent of value-adding supported by a journal. Stars should not be jeopardised by an instance of fraud, but should be won or lost on the procedures in place to detect and deal with fraud and other misdemeanours. Thomson Reuters could show leadership with such a new certification system, or failing them editors should collaborate to achieve the same end. There are useful lessons from the 20 years of experience with forest certification (e.g., Vlosky & Ozanne 1997, Cashore et al 2005), where independent scrutiny has been desirable to motivate progressive improvement in forest management. Forestry experience reveals the importance of a well-resourced executive group overseen by a strong board representing stakeholder interests. Thomson Reuters or other providers seeking to fill this space would do well to take inspiration from the forest certification experience.

## Conclusion

The Thomson Reuters impact factor (TRIF) suffers so many weaknesses, that a major overhaul is warranted, and journal editors and other users should cease using the TRIF until Thomson Reuters has addressed these weaknesses. Urgent improvements include the adoption of a ‘like-with-like’ basis (i.e., citations to articles, divided by the count of articles only), the use of verified one-to-one links only (this would unite authors, editors and Thomson Reuters in quality control); the adoption of a more appropriate reference interval (the present two year interval is too short for many disciplines), the introduction of confidence intervals, and the rounding of reported indices to a more appropriate number of digits. Failing action by Thomson Reuters, journal editors should collaborate as they have come with COPE to introduce a journal certification system that acknowledges procedures to maintain quality: procedures that add value and restrict plagiarism and fraud. The future of quality science communication lies in the hands of editors.